\documentclass[]{elsarticle}

\usepackage{amssymb, amsmath,graphicx}

\begin{document}

\begin{frontmatter}

\title{Gravitation at the Josephson Junction}

\author{Victor Atanasov}
\address{Department of Condensed Matter Physics, Sofia University, 5 boul. J. Bourchier, 1164 Sofia, Bulgaria}
\ead{vatanaso@phys.uni-sofia.bg}

\begin{abstract}
A geometric potential from the kinetic term of a constrained to a curved hyper-plane of space-time quantum superconducting condensate is derived. An energy conservation relation involving the geometric field at every material point in the superconductor is demonstrated. At a Josephson junction the energy conservation relation implies the possibility to transform electric energy into geometric field energy, that is curvature of space-time. Experimental procedures to verify that the Josephson junction can act as a voltage-to-curvature converter are discussed.
\end{abstract}


\end{frontmatter}

The success of the Laser Interferometer Gravitational Wave Observatory (LIGO) and its sister collaboration, VIRGO\cite{LIGO}, in observing the geometric field's ripples in space-time has in addition to opening a new experimental method of astrophysical observation, proved a practical scheme for measuring the space-time metric, that is LIGO had seen the death spiral of a pair of  black holes, through the dislocation and the associated change in optical (and physical) path travelled by light beams in the arms of an interferometer  as the gravitational perturbation travels between the reflection points. In theory, the geometry of space-time, that is the components of the Riemann curvature tensor can be reconstructed by taking measurements of the deviation of two adjacent light paths (geodesics)\cite{GeodDev}.

Having in mind this spectacular success in experimental astrophysics, we pose the following questions: {\it i.) Are there other experimentally relevant strategies for detecting space-time geometry? ii.) Is there other extremely sensitive measurement technique besides two light beam interference that can potentially sense the geometric field?} The present paper aims at answering both questions and is organised as follows. The first section discusses the appearance of a geometric potential term from the kinetic energy term in the Schr\"odinger equation. In order to improve on the detection possibilities we assume this constrained to a hyper-plane quantum dynamics to concern a superconducting condensate. The second section focuses on the hydrodynamic interpretation of the governing quantum equation and reveals that the geometric field enters it on an equal footing with the "quantum potential" (in Bohm's views) thus making its way as a real force moving the condensate superfluid. The third section contains the proof that the emergent geometric field enters an energy conservation relation valid at each point of the superconducting condensate. Based on this conservation relation applied to a Josephson junction, an experimentally verifiable voltage-to-curvature conversion effect is proposed in the fourth section. The fifth section discusses possible experimental methodologies to test the reality of the effect. 

\section{The geometric field in the Schr\"odinger equation}

The general theory of relativity conveyed an understanding of phenomena such as the distortion of time-space by a gravitational or acceleration field\cite{AE}. However, the manner in which the curvature of space-time, that is the Riemannian space, affects the electronic properties of condensed matters systems on a microscopic scale is largely unknown and due to its experimental accessibility of great interest\cite{C60}. 

Riemannian geometric effects in a quantum system, which can either be free or constrained, stem from the dependance of the kinetic term on the metric of the embedding space or the metric of the sub-manifold onto which the quantum system is constrained by a confining potential (rigid chemical bond; electrostatic attraction).

The problem of constraining particle motion to a curved sub-manifold embedded in a Euclidean space $\mathbb{R}^n$ can be resolved in one of two alternative ways: 
i.) In the {\it intrinsic} quantization approach, the motion is constrained to the curved sub-manifold in the first place. A  Hamiltonian is constructed from generalized coordinates and momenta intrinsic to the sub-manifold and the system is quantized canonically. As a result, the embedding space is inaccessible and the quantum system depends only on the geometry intrinsic to the sub-manifold/hyper-plane\cite{Leaf, deWitt}; 
ii.) In the {\it confining} potential approach, a free in the embedding space quantum particle
is subjected by a normal to the sub-manifold force that in effect confines the dynamics onto it. The effective Hamiltonian depends on the intrinsic geometry and on the way this sub-manifold
is immersed in the embedding space. 

On one hand, the intrinsic quantization procedure is plagued with ordering ambiguities
that allow for multiple consistent quantization procedures different by a term
proportional to the curvature of the sub-manifold\cite{deWitt, Birrell&Davies, Wald}. 
On the other hand, the confining potential procedure leads to a unique effective Hamiltonian that depends on the constraint. In real microscopic quantum systems,
constrained motion is a result of a strong confining force (electrostatic; rigid chemical bonds, ect.). Therefore, confining potential formalism seems a physically more realistic approach to constraints \cite{Jensen&Koppe, daCost*81, ogawa, Goldstone&Jaffe, Exner*01, Schuster&Jaffe}.

The non-relativistic quantum mechanics in a three dimensional hyper-plane of the four dimensional curved space-time can be treated in a well established manner\cite{Wald}. In this case the embedding space-time is non-Euclidean but equipped with a metric. This four-dimensional metric is related to the matter distribution by the Einstein equation. Suppose the four dimensional space-time $M$ is topologically the product $M\cong \mathbb{R}\times \Sigma,$ where $\Sigma$ represents a space-like three dimensional hyper-plane. We can then foliate $M$ by a one parameter family of imbeddings given by the map $\tau_t:$ $\Sigma \to M$ such that $\Sigma_t=\tau_t(\Sigma) \subset M,$ that is $\Sigma_t$ is the image of the map $\tau$ in $M$ for a fixed ''time'' $t.$ We assume that the leaves $\Sigma_t$ are space-like with respect to the metric in $M.$ As a result, there exists in $M$ a time-like field normal to the leaves $\Sigma_t,$ therefore there is a notion of future and past. This time evolution vector field $t^a=(\partial / \partial t)^a,$ satisfies $t^a \nabla_a t=1,$ so that local coordinates $t, x^1, x^2, x^3$ (satisfying $t^a \nabla_a x^b=0,$ for $b=1,2,3$) can be introduced. In effect, the space-time is splittable into 3+1 dimensions and the induced Riemannian metric $g_{ij}$ onto the three dimensional $\Sigma_t$ can be used to write the Laplace-Beltrami operator $\Delta_{LB}$, which is the kinetic energy term in the Schr\"odinger equation for the subjected to the geometric field quantum particle or quantum condensate
\begin{equation}
\Delta_{LB} \Psi = \frac{1}{\sqrt{|g|}} \partial_{j} \left( \sqrt{|g|} g^{jk} \partial_{k} \Psi  \right)=g^{jk}\partial_j \partial_k \Psi - g^{jk} \Gamma^{l}_{jk} \partial_l \Psi. 
\end{equation}

The emergence of the geometric field from the kinetic term can be made clearer in the vicinity of the origin where the following Taylor expansion of the induced metric in normal coordinates applies: $g_{ij}=\delta_{ij}-\frac13 R_{ikjl}x^{k} x^{l} + O(|x|^3)$ \cite{Viaclovsky} and expanding the square root of the determinant of the metric yields:
\begin{equation}\label{eq:g}
\sqrt{|g|}= 1 - \frac16 R_{jk}x^{j} x^{k} + O(|x|^3).
\end{equation}
Using a standard re-normalisation of the wave-function $\Psi=\psi/|g|^{1/4}$ and keeping the lowest order terms (the only relevant for the quantum dynamics) in the Taylor expansion we get for the kinetic term in the Schr\"odinger equation
\begin{eqnarray}
\nonumber -\frac{\hbar^2}{2m}\Delta_{LB} \frac{\psi}{|g|^{1/4}} && =  \frac{1}{|g|^{1/4}} \left( -\frac{\hbar^2}{2m}\Delta \psi + \frac{\hbar^2}{4m} \frac{g^{lk} \partial_{l} \partial_{k} \sqrt{|g|}}{\sqrt{|g|}} \psi  \right) + O(|x|)\\
&& =  \frac{1}{|g|^{1/4}} \left( -\frac{\hbar^2}{2m}\Delta \psi - \frac{\hbar^2}{24m}R \psi  \right) + O(|x|). 
\end{eqnarray}
Here $\Delta$ is the Laplacian on flat space.
Adding an additional potential $U(x^1, x^2, x^3) $ that may act in the system we convey the complete symbolic equation with which we will further work with
\begin{equation}
-\frac{\hbar^2}{2m}\Delta \psi  + \left( V_{Geom} + U \right) \psi=i\hbar \partial_t \psi. 
\end{equation}
Here
\begin{equation}
V_{Geom}=- \frac{\hbar^2}{2m} \alpha R, 
\end{equation}
where $R$ is the three-dimensional Ricci scalar curvature and $\alpha=1/12$ is a numeric coefficient. The emergence of a geometric potential from the kinetic term is obvious. Such a term is a standard coupling term between curvature and a quantum field in quantum field theory in curved space-time. 

Note, the particular form of the geometric potential may vary and in the case of a constraining potential approach takes the expressions: i.) $V_{Geom}=- \frac{\hbar^2}{8m} \kappa^2,$ where $\kappa$ is the principle curvature of a space curve embedded in $\mathbb{R}^3$ \cite{daCost*81, Goldstone&Jaffe}; ii.) $V_{Geom}=- \frac{\hbar^2}{8m} (\kappa_1 - \kappa_2)^2,$ where $\kappa_i,$ for $i=1,2$ are the principle curvatures of a surface embedded in $\mathbb{R}^3$ \cite{Jensen&Koppe, daCost*81};  iii.) $V_{Geom}=- \frac{\hbar^2}{8m} \left[ \kappa_3 \left(\kappa_3 - 2(\kappa_1+\kappa_2) \right)  \right. +$ $ \left.  (\kappa_1 - \kappa_2)^2 \right],$ where $\kappa_i,$ for $i=1,2,3$ are the principle curvatures of a three dimensional manifold embedded in $\mathbb{R}^4$\cite{Schuster&Jaffe}.

When electric field (defined with the potential $V$) and magnetic field, defined through the vector potential $\vec{A},$ are present the Schr\"odinger equation takes the following form
\begin{equation}\label{Schrodinger&R}
\frac{1}{2m}\left(\frac{\hbar}{i} \nabla - q \vec{A}  \right).\left(\frac{\hbar}{i} \nabla - q \vec{A}  \right) \psi + qV\psi+ \left( V_{Geom} + U \right) \psi=i\hbar \partial_t \psi. 
\end{equation}

\section{Hydrodynamic interpretation of the condensate wavefunction}

Suppose, that we deal with a Cooper pair condensate inside a superconductor. The Schr\"odinger equation for the Cooper pair will be the above equation (\ref{Schrodinger&R}) with $q=2e,$ that is twice the charge of the electron. This equation will describe the state of the entire condensate. Therefore, we may write  $\psi=\sqrt{\rho(\vec{r})}e^{i\theta(\vec{r})},$ where $\rho(\vec{r})$ is the the charge density of the condensate and $\theta(\vec{r})$ its phase. Upon substitution of this form of the wave-function into (\ref{Schrodinger&R}) we can separate the real and imaginary part of the equation to arrive at slightly modified standard result:
\begin{eqnarray}\label{eq:rho_t}
\frac{\partial \rho}{\partial t}= - \nabla. \vec{J}, \quad \vec{J}=\vec{v}\rho=\frac{1}{m}\left( \hbar \nabla \theta - q \vec{A} \right) \rho\\\label{eq:theta_t}
\hbar \frac{\partial \theta}{\partial t}=-qV-\frac{1}{2m} \left( \hbar \nabla \theta - q \vec{A} \right)^2 + \frac{\hbar^2}{2m} \left( \frac{\Delta \sqrt{\rho}}{\sqrt{\rho}} + \alpha R \right)
\end{eqnarray}
Here $\vec{J}$ is the current density, which in the case of a superconducting condensate stands also for the probability current. The generalised momentum is contained in the expression for$\vec{J}:$ $\vec{p}= \hbar \nabla \theta - q \vec{A},$ therefore the current density is just the velocity of the superconducting current times the charge density.

Taking the gradient of the whole equation (\ref{eq:theta_t}) and expressing $\nabla \theta$ from (\ref{eq:rho_t}) (akin to \cite{Feynman}) we obtain the modified version of the hydrodynamic interpretation of the quantum condensate dynamics:
\begin{eqnarray}\label{eq:dv/dt}
\frac{d \vec{v}}{dt}=\frac{\partial \vec{v}}{\partial t} + \vec{v}.\nabla \vec{v}=\frac{1}{m} \vec{F}_{L} + \frac{1}{m}\nabla \left[ \frac{\hbar^2}{2m} \left( \frac{\Delta \sqrt{\rho}}{\sqrt{\rho}} + \alpha R \right) \right],\\
\nabla \times \vec{v}=-\frac{q}{m}\vec{B}
\end{eqnarray}
where $\vec{F}_{L}=q\vec{E} + q \vec{v} \times \vec{B}$ is the Lorentz force acting on the charged Cooper pairs. These two equations are the equations of motion of the superconducting Cooper pair fluid in the presence of an induced from the embedding space-time curvature, which in this paper ofter refer to geometric field. Note, the geometric field enters the gradient of the mystical quantum mechanical potential, recognised by Bohm as a unique interaction with the $\psi-$field itself\cite{Bohm}.  As a result, the geometric field and the $\psi-$field (in view of Bohmian quantum mechanics) have similar and competing action. 

Next we recall the London equations for the quantum superconducting current density\cite{London}
\begin{equation}\label{eq:london}
\vec{J}= \hat{\Pi} \vec{A},
\end{equation}
where $\hat{\Pi}=-\rho q/m$ (see eq.(21.20) in \cite{Feynman})  For brevity we will call the introduced quantity $\hat{\Pi},$ which can be either of scalar or tensorial character, the polarisation operator. A correct microscopic theory of superconductivity can produce an expression for it in therms of the energy gap and critical temperature\cite{BCS}.

Note, an important issue needs to be addressed, namely to what extent the London equations hold in curved space-time. The above London equation is implicitly contained in the Schr\"odinger equation within the form of the canonical momentum. However, in the curved space case the canonical momentum (\ref{eq:rho_t}) coincides with the flat space case (see \cite{Feynman}), therefore we will not seek any generalisation of (\ref{eq:london}). An additional reinforcement of this choice comes from the original London brothers' derivation, namely the super-current is being accelerated under the influence of external electromagnetic fields as if made up of free charged particles. Therefore in curved space-time $J^{\nu}=\partial_{\mu} \sqrt{-g} F^{\mu \nu},$ where  $F^{\mu \nu }$ is the electromagnetic tensor \cite{LL}. Reducing the above to the space part and using (\ref{eq:g}) within the zero-th order in the vicinity of the origin (the same approximation as the one used in the derivation of (\ref{Schrodinger&R})) we obtain $J^{\nu}=\partial_{\mu} (1 + O(x^2) ) F^{\mu \nu} \approx\partial_{\mu} F^{\mu \nu},$ which coincides with the flat space case, therefore the second London equation (produced by taking a curl from this one) should also coincide with the flat space one, that is (\ref{eq:london}).  London equations are grounded in the electrodynamics of the superconductor and more specifically the phenomenological description of its ideal diamagnetism. We do not have any indication that this material property is rendered invalid in curved space-time.

In addition, the London theory can be viewed as a limit (the London limit) of the phenomenological Ginzburg-Landau theory, which in the case of curved  space-time is extended with an extra term encoding the interaction with the geometric field, besides the standard extension of the covariant derivatives to include the Christoffel symbols\cite{victor}.  The supercurrent operator emerging from this 
approach coincides with (\ref{eq:rho_t})\cite{victor1}, therefore the above conclusion on the validity of the London theory in curved space-time is preserved. London theory remains valid also in the case of the gravito-electromagnetic approximation to the Einstein field equations\cite{gravitoEM}.

Next, we divide both sides of (\ref{eq:london}) by the current density $\rho$ and then differentiate with respect to time 
\begin{equation}\label{eq:london/rho}
\frac{d \vec{v}}{d t} = \frac{d }{d t} \frac{\vec{J}}{\rho}=\frac{d }{d t} \frac{\hat{\Pi} \vec{A}}{\rho},
\end{equation}
only to equate the r.h.s. of (\ref{eq:dv/dt}) with the r.h.s of (\ref{eq:london/rho})
\begin{equation}
\vec{F}_{L} + \nabla \left[ \frac{\hbar^2}{2m} \left( \frac{\Delta \sqrt{\rho}}{\sqrt{\rho}} + \alpha R \right) \right]=\frac{d }{d t} m \frac{\hat{\Pi} \vec{A}}{\rho}.
\end{equation}

\section{The geometric effect}

In the case when the superconducting state is robust, we may assume that 
\begin{equation}
\frac{d \hat{\Pi}}{d t} \approx 0, \quad  \frac{d \rho}{d t} \approx 0 \quad \& \quad \Delta \sqrt{\rho}\approx 0
\end{equation}
the current density in the superconductor is approximately constant as well as the polarisation operator (no internal changes in the microscopic mechanism). The ideal diamagnetism of the superconducting state reduces the Lorentz force to its electrostatic part, which is non-vanishing only in the case when a Josephson junction is present (two separated conducting domains at different electrostatic potentials). Finally, the above simplifications yield
\begin{equation}\label{eq:nabla_R}
\nabla \left[ \frac{\hbar^2}{2m}  \alpha R \right]= - q\vec{E} +  m \frac{\hat{\Pi} }{\rho} \frac{d \vec{A}}{d t}.
\end{equation}
which is simply an expression for the conservation of energy. 

Let us take a line integral of the above along an open path from point A to point B
\begin{equation}\label{eq:nabla_R_2}
\frac{\hbar^2}{2m}  \alpha \int_A^B \nabla R . d\vec{l} = - \int_A^B q\vec{E} . d\vec{l} +  m \frac{\hat{\Pi} }{\rho} \int_A^B \frac{d \vec{A}}{d t} . d\vec{l}.
\end{equation}
Next we introduce the geometric field energy
\begin{equation}
W_g(\vec{r})=\alpha \frac{\hbar^2}{2m}  R (\vec{r}),
\end{equation}
next recall that $E_{ind}=-\partial \vec{A} / \partial t,$ that is $q\int_A^B \vec{E}_{ind} . d\vec{l}=\mathcal{E}_{ind}(B) - \mathcal{E}_{ind}(A)$ is the electromotive potential difference between the two points 
and
$$ 
\frac{d \vec{A}}{d t}=\frac{\partial \vec{A}}{\partial t} + \vec{v}.\nabla \vec{A} \quad
\frac{d A_i}{d t}=\frac{\partial A_i}{\partial t} + \frac{\partial r_j}{\partial t} \frac{\partial A_i}{\partial r_j}. 
$$
Finally (\ref{eq:nabla_R_2}) can be re-written using (\ref{eq:rho_t}) as 
\begin{eqnarray}
W_g(B) - W_g(A) &=& q \left[U_{stat}(B) - U_{stat}(A) \right] \\
\nonumber && -\frac{m}{q}\frac{\hat{\Pi} }{\rho} \left[ \mathcal{E}_{ind}(B) - \mathcal{E}_{ind}(A) \right]+  \frac{m}{q} \frac{\hat{\Pi} }{\rho} q \int_A^B ( \vec{v}.\nabla \vec{A} \,) . d\vec{l}
\end{eqnarray}
Now, suppose the Cooper pair charge velocity is constant at the two adjacent points, than the last integral quantity measures the difference in the interaction energy $\delta W_{int}$ between the Cooper pairs  and the vector potential at the two points:
\begin{eqnarray}\label{eq:interaction}
q\int_A^B ( \vec{v}.\nabla \vec{A} \,) . d\vec{l} =q \vec{v}.\vec{A}(B) - q\vec{v}.\vec{A}(A) = \delta W_{int}
\end{eqnarray}
Introducing the electrostatic energy $W_{stat}=qU_{stat}(\vec{r})$ we can put (\ref{eq:nabla_R_2}) in its final form
\begin{eqnarray}
\nonumber && W_g(B) -W_{stat}(B) + \frac{m}{q} \frac{\hat{\Pi} }{\rho} \left[ \mathcal{E}_{ind}(B) - W_{int}(B)\right] \\ 
 && \qquad =W_g(A) - W_{stat}(A) +\frac{m}{q}\frac{\hat{\Pi} }{\rho} \left[  \mathcal{E}_{ind}(A) -  W_{int}(A)   \right].
\end{eqnarray}
As a result of introducing the scalar polarisation operator from (\ref{eq:london}) and \cite{L9}, the following conserved quantity at each material point of the superconductor emerges:
\begin{eqnarray}\label{eq:conservation}
W_g(\vec{r}) - \mathcal{E}(\vec{r}) +  W_{int}(\vec{r}) ={\rm const}.
\end{eqnarray}
Here $\mathcal{E}(\vec{r})=W_{stat}(\vec{r}) - \mathcal{E}_{ind}(\vec{r})$ is the electrical energy of the Cooper pairs.

\section{Direct and reversed effect}

In the previous section we have seen that the geometric field is equivalent to an electric field in the superconductor (\ref{eq:nabla_R}) which statement is analogous to the law of conservation of energy (\ref{eq:conservation}). Therefore, provided the superconducting element is homogeneous, we can expect $W_{int}(\vec{r})={\rm const}$ and as a result of the perfect conductor aspect of the superconducting state, we can also expect a redistribution of the shifted by the geometric field charges inside the superconductor in order to maintain the superconductor at a constant potential. 

A completely different behaviour can be expected at the Josephson junction. We will discuss two cases of the junction, one between superconducting sides made from the same superconductor (symmetric) and one between two different superconductors (asymmetric). 

Clearly, in the symmetric junction, the interaction energy $W_{int}(\vec{r})$ will be the same on both sides. However the electrostatic potential on the two sides can be different and the voltage drop $U$ can be equated to the geometric potential, that is the curvature scalar R itself. In effect, the difference in the geometric field between the two sides $\delta R$ can produce a voltage drop at the junction (direct effect) or the voltage drop across the junction can produce curvature difference (reversed effect):
\begin{eqnarray}\label{eq:direct/reversed W=0}
\alpha \frac{\hbar^2}{2m}  \delta R  = qU.
\end{eqnarray}
In this case, we may regard the Josephson junction as a curvature-to-voltage converter with the following ratio ($m$ is the free electron mass):
\begin{eqnarray}\label{eq:conversion factor}
 1[V] \approx 6.3 \times 10^{20}  [{\rm m}^{-2}].
\end{eqnarray}
Note, the only difference the asymmetric junction can introduce is the difference in the interaction energy $\delta W_{int}$ between the supercurrent and the electromagnetic field at the two sides of the junction. Suppose the supercurrent  flows in the junction at vanishing potential difference $U \to 0$, then the interaction energy gradient can produce rippling in the geometric field according to:
\begin{eqnarray}\label{eq:conservation}
\alpha \frac{\hbar^2}{2m}  \delta R   \approx - \delta W_{int}.
\end{eqnarray}
The interaction energy (\ref{eq:interaction}) is a function of the supercurrent drift velocity and one may view the asymmetric effect as  produced by a sharp change in momentum, which converts into rippling of the geometric field. We may expect a back-reaction on the entire junction as well. The kinetic energy of the bulk material can certainly be included in (\ref{eq:conservation}). The back reaction will increase with the increase in the difference between the interaction energy of the supercurrent with the electromagnetic field on both sides.

\section{Proposed experimental verification}

An argument that the Josephson junction can act as a reversible curvature-to-voltage converter was presented in the previous section. In effect, the argument is prone to experimental testing and now we will discuss how and to what extent. Note, the conversion factor (\ref{eq:conservation}) points to the impossibility to observe the travelling ripples in space-time, that is gravitational waves, with a Josephson junction. Along the span of the junction (few angstrom [$\AA$]) the expected difference in the induced scalar curvature is very small $\delta R > 10^{-21}$[m$^{-2}$], therefore according to (\ref{eq:conservation}) we may not hope for potential difference greater than $10^{-40}$[V] which is unmeasurable. We are in a position to answer the questions from the introduction. We are unable to sense the geometric field produced by a gravitational wave at the Josephson junction.

However, since we expect that the effect is reversible, we may attempt to create a geometric field at the junction by an electric discharge between its sides. The greater the potential difference that can be created between the sides, the greater the geometric field that could be created. We are unaware of the dynamics of the created geometric field and do not have any governing equations at the present stage of discussion on its propagation. Nevertheless, we can propose two detection methods which in theory can confirm the proposed effect.

\begin{figure}[h]
\begin{center}
\includegraphics[scale=0.35]{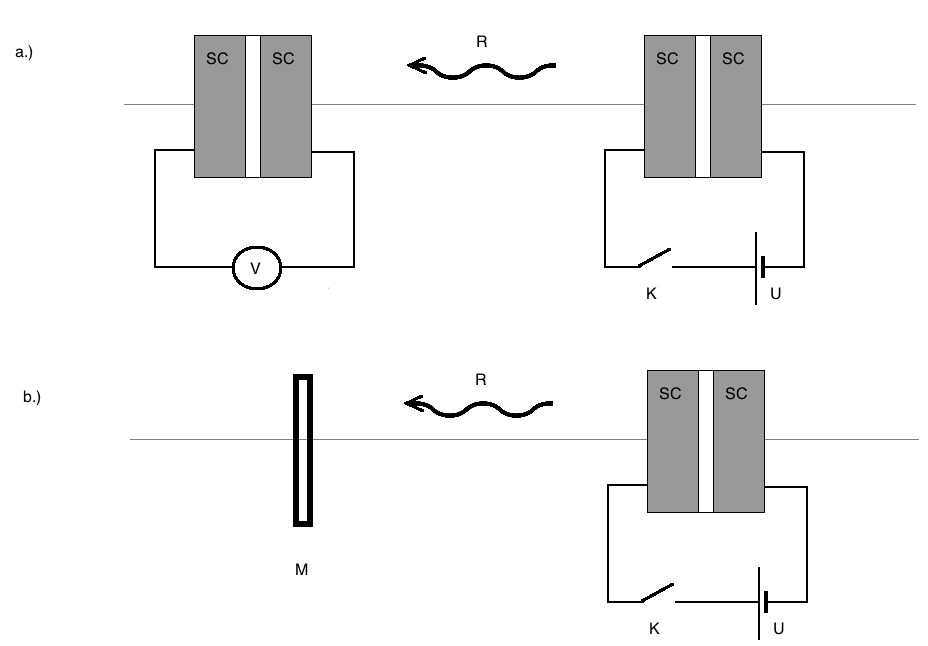}
\caption{\label{figure}Detection schemes: a.) Detection based on the reversibility of the effect. Here a discharge with e.m.f. U is conducted in one of the junctions as the circuit is closed via key K. The emitted geometric field R is detected with the second Josephson junction with the voltage induced between its sides as the geometric pulse impinge on it; b.) Detection based on the dislocation induced by the geometric pulse R on a mirror M part of an interferometer measurement circuit.}
\end{center}
\end{figure}

The first approach stems from the reversibility argument, see Figure \ref{figure}(a). Suppose we have two junctions in close proximity. We have no firm reason to choose a particular set-up, but choose to discuss the idea of the experiment with the two junctions placed along a line in such a way that the plane of the junctions (the insulating layer) is normal to the imaginary line connecting them. Both junctions should be magnetically and electrically shielded from each other and the surrounding environment. One of them will serve as an emitter and the other as detector. A high-voltage discharge should be conducted in the emitter and an induced voltage drop should be recorded at the detector. Provided such an electric potential difference is recorded via proper coincidence scheme (an additional detector junction involved), we may confirm that the geometric effect at the Josephson junction is a physical reality.

The second approach involves the ability of the geometric field to impart motion to objects with inertia, see Figure \ref{figure}(b). The geometric field while a measure of the curvature of space-time is an acceleration field as well, or better induces force (in the lab frame) on a free object of inertia via the Newton's second law $\vec{F}=m\vec{a},$ where $\vec{a}$ is the imparted acceleration. We can give a rough  estimate of the imparted acceleration in order to come up with an experimental procedure to verify the effect. There are two possibilities to arrive at an estimate. The first approach involves the use of the Gauss's law for gravitation $\nabla.\vec{a}=-4 \pi G \rho_{matter}$ and the $(t,t)$ component of Einstein's field equations for a perfect fluid (where $T_{tt}=-\rho_{matter} c^2$): $G_{tt}=8 \pi G \rho_{matter}/c^2.$ Here $c$ is the velocity of light in vacuum, $G$ is Newton's gravitation constant and  $\rho_{matter}$ is the matter density. Next we make use of an exact result valid for a 3+1 decomposition of space-time:$G_{tt}=R/2,$ where $R$ is the scalar Ricci curvature of the three-dimensional hyper-surface\cite{Giulini}. Combining these relations we end up with $\nabla . \vec{a} = R c^2 /4,$ that is equivalent to 
\begin{equation}
a=\frac{R c^2}{4} \delta x
\end{equation}
in the case when the imparted acceleration is in one dimension along the span of $\delta x.$ Interestingly, a similar result is obtained if one takes up the geodesic deviation equation $D^2 x^{\mu}/d\tau^2 = R^{\mu}_{\; i j \nu} T^{i} T^{j} x^{\nu},$ where $T^{i}$ is the 4-velocity of an object travelling along the geodesic and $x^{\mu}$ is the deviation vector. Note, the Riemann tensor enters the relation directly. The geodesic deviation measures the acceleration with which two neighbouring geodesics deviate from each other in the curved geometry. Suppose $T^{i}$ is a unit vector in the time direction and $x^{\nu}=x_0^{\nu}+ a^{\mu} t^2/2 + O(t^3),$ where $x_0^{\nu}$ is a constant, then the geodesic deviation equation in one dimension is approximately
$a \approx R c^2 \delta x$ upon a substitution of the Riemann tensor component with the Ricci scalar curvature. Most importantly, the two expressions agree in the order of magnitude and the expected acceleration a geometric field pulse with a magnitude of $R\sim 10^{20}  [{\rm m}^{-2}]$ and a width of the size of the Josephson junction $10^{-10} [{\rm m}]$ can impart on an object along its path is enormous $a \sim 10^{25}g.$ The dislocation such a pulse can cause is proportional to the time of its duration squared. However, we have no estimate of this quantity, but given the large value for the acceleration even a femtosecond pulse can lead to substantial dislocation of the order of meters. We also have no estimate of the spread with distance of this geometric field pulse and believe the assumed value for the scalar curvature and imparted to a material body acceleration to be largely exaggerated.

Now suppose we conduct a high-voltage discharge in a Josephson junction and try to measure the dislocation of a mirror, mechanically shielded from the junction. Such a dislocation will be induced by the emitted at the junction geometric field. The position of the mirror with respect to the junction is unknown, therefore few geometric set-ups should be tried. Next, in order to increase the sensitivity of the experiment, we suggest the inclusion of the detection mirror as a primary or secondary mirror  in a Michelson interferometer. Provided a dislocation in the mirror is recorded we may confirm the generation of a geometric field in a Josephson junction.

\section{Conclusion}

In conclusion we would like to point out the origin of the  geometric potential from the kinetic term of a constrained to a curved three-dimensional hyper-plane of space-time quantum mechanical condensate (superconductor). This potential makes its way into the hydrodynamic interpretation of the Schr\"odinger equation and enters it on an equal footing with the Bohm's "quantum potential." When external electromagnetic field is included in the dynamics and a suitable simplifications applied, one is able to derive an obvious energy conservation relation at every material point in the superconductor. This conservation relation includes a geometric field part associated with the curvature of the hyper-plane. It turns out that at a tunnelling junction (Josephson junction) the energy conservation relation implies the possibility to transform electric energy into geometric energy, that is create curvature in the hyper-plane and vice versa. In effect, it turns out that the Josephson junction can act as a voltage-to-curvature converter. Experimental procedures are discussed in hope the present study invites experimental effort to verify the effect.

The author declares that there is no conflict of interest regarding the publication of this paper.

\end{document}